# Topology-guided vortices in a polariton condensate

Andrea Zacheo[1,2], Marco Marangi[1,2], Nilo Mata-Cervera[1,2], Yijie Shen[1,2,3], Giorgio Adamo[1,2], Cesare Soci[1,2,3,*]

[1] *Centre for Disruptive Photonic Technologies, The Photonics Institute, Nanyang Technological University, Singapore 637371*

[2] *Division of Physics and Applied Physics, School of Physical and Mathematical Sciences, Nanyang Technological University, Singapore 637371*

[3] *School of Electrical and Electronic Engineering, Nanyang Technological University, Singapore 639798, Singapore*

**Email : csoci@ntu.edu.sg*

**A major challenge in polariton fluids is achieving deterministic control over the spin texture of the macroscopic condensate wavefunction, which dictates the nucleation and dynamics of topological excitations, such as vortices, solitons, and strings. Existing approaches typically rely on external gauge fields to indirectly access the polariton pseudospin, resulting in configurations that are weakly constrained by the cavity modes and therefore highly sensitive to disorder and fluctuations. Here, we report the generation of spin polaritons constrained to the topology of a bound state in the continuum (BIC) metasurface with broken inversion symmetry carved into a polycrystalline halide-perovskite film. Geometry-induced polariton condensation under spin-momentum locking gives rise to a pair of half-vortices of opposite spin, intrinsically pinned to polarization strings, emerging as topological extensions of the vortex cores. Consequently, varying the excitation density drives a controlled displacement of the half-vortices along the trajectories imposed by the strings, hindering their mutual annihilation across an interposed topological domain wall. This approach establishes cavity geometry as an intrinsic source of spin textures to guide vortex displacement in driven quantum fluids, opening a route toward the generation of robust topological excitations within structurally disordered materials.**

**Keywords:** *exciton-polariton condensate, half-vortices, spin-momentum locking, monolithic perovskite metasurface, bound state in the continuum (BIC), topology engineering, spin-orbit interactions, polariton spin texture, dissipative-driven quantum fluid.*

## Introduction

Vortices and topological charges have often been used as a unifying concept across physics, spanning from subatomic particles[1] and quantum fluids[2–4] to non-linear optics[5] and cosmology[6]. Among the defining features of vortices, characterized by an integer multiple of 2π phase winding of the wavefunction across a point-like singularity[7], are the circulation of the energy flow around the core, where the density of particles reaches a minimum and the rotational velocity formally diverges to infinite, and the presence of a topological charge whose sign is determined by the direction of circulation[8]. While vortices provide the most immediate and localized manifestation of topological order, they belong to a broader hierarchy of composite topological objects.[9] These include extended defects such as strings, one-dimensional manifolds where the order parameter vanishes or undergoes a discontinuity,[10,11] and walls consisting of boundaries that separate domains with different symmetry configurations or topological invariants.[12–14] Typically these defects do not appear in isolation but as intrinsically interconnected structures, with mutual arrangement and interactions governed by the topology of the order parameter field.[15,16] Topological defects have been extensively investigated in polariton superfluids, typically realized in excitonic quantum wells embedded in microcavities. Fundamental studies have focused on vortices within the "hydrodynamic" approximation such as full vortices[17], half vortices[18,19], spin vortices[20,21], and on fluctuation-generated vortices like the ones observed during Berezinsky-Kosterlitz-Thouless (BKT) transition[22]. In this context, the pseudospin, defined as a coordinate on the Poincaré (or Bloch) sphere, provides an effective representation of the polariton spin state, offering a natural framework to describe the formation and dynamics of emerging topological defects[23]. The generation and control of spin-polarized polariton condensates in microcavity has been achieved by employing either inhomogeneous[24] or structured light sources, which entail topological singularities[25,26], by integrating spin-selective optical elements like nematic crystals[27–30], by employing an external magnetic field[31], and by exploiting the TE-TM splitting of microcavities[32]. In addition, previous studies on the formation and dynamics of vortices in microcavity polariton condensates have shown indirect and stochastic control over vortices nucleation and spatiotemporal evolution, that were found to follow complex and irregular trajectories[33–35]. To date, control over vortex formation, positioning, and interaction remains a significant challenge in polariton systems.

Recently, photonic metasurfaces supporting optical bound states in the continuum (BICs) have emerged as a versatile open cavity architecture to enter the strong-coupling regime[36] and induce exciton-polariton condensation[37]. BICs are non-radiative states in the continuum spectrum protected by lattice symmetry, which manifest as dark modes at the Γ-point of the reciprocal space[38]. Due to the symmetry mismatch between the photonic mode profiles and the out-of-plane propagating waves, BICs are fully decoupled from the far field radiation and therefore cannot be accessed by external excitation[39]. This results in strong photon confinement in the near field and a theoretically infinite Q-factor of the cavity mode. Beyond this conventional symmetry-based picture, BICs can also be understood from a topological perspective: in momentum space, the far-field polarization winds around the BIC, forming a vortex whose charge is robust against continuous perturbations. As a result, optical BICs carry an invariant integer topological charge, defined by the winding number of the polarization director, analogous to quantized vortices in quantum fluids[40]. When the in-plane symmetry is broken, these ideal BICs acquire a finite radiative linewidth, and their resonance Q-factor is correspondingly reduced. The resulting modes, now weakly coupled to the far-field, are referred to as *quasi*-BICs (*q*BICs). Importantly, the nature of symmetry-breaking also determines the spin texture of the radiated field. With appropriately engineered perturbations, *q*BICs can exhibit spin-selective emission and momentum-dependent splitting of their dispersion, an optical analogue of the electronic Rashba effect[41]. These features become particularly compelling for BIC metasurfaces fabricated directly in strongly excitonic media where the spatial and energetic overlap between photons and excitons resonances is indistinguishable. In this case, the metasurface geometry not only governs the structure of the polaritonic modes but also provides selective control over the polariton pseudospin through precise design of the underlying symmetry. Among the available material platforms, halide perovskites (hybrid organic-inorganic semiconductors with strong exciton binding energy, high luminescence, and tunable bandgap) have emerged as ideal candidates for the realization of both optically[42] and electrically[43] driven monolithic BIC metasurfaces. In these systems, exciton-polariton formation, condensation and emission carrying spin and topological charges have all been recently demonstrated.[44–46]

Here, we exploit a $q$BIC monolithic metasurface with in-plane broken inversion symmetry carved into a polycrystalline halide-perovskite film to directly access and control of the intrinsic polariton pseudospin, as evidenced by the appearance of a stable half vortex-antivortex (V-AV) pair in the condensate emission, each carrying opposite topological charge. Excitation density dependent measurements unveil a progressive displacement of the V-AV positions, reflecting a density-driven spatial evolution that is constrained by the intrinsic polarization texture of the condensate rather than by external control parameters. The observed trajectories originate from the emergence of polarization strings attached to the vortex cores, generated by the symmetry breaking that induces an anisotropic spatial polarization distribution analogous to strings appearing in atomic superfluids[47]. Acting as geometrical guides, these strings delineate the paths along which the half-vortices can evolve, thereby guiding their motion, while the emergence of a topological domain wall located at the centre of the condensate prevents their mutual annihilation. Despite the inherent structural disorder of the polycrystalline films arising from grain boundaries and structural defects, the topological robustness provided by the metasurface design preserves the condensate coherence and protects vortex motion in the fluid. Taken together, these results define a qualitatively new framework in which topology, geometry, and nonlinear polariton physics are intrinsically coupled to govern vortex arrangement and interactions. By achieving this level of control within a scalable perovskite platform, the present work advances both the conceptual understanding and practical implementation of robust topological excitations in polariton fluids.

**Symmetry breaking in *q*BIC metasurface induces spin-orbit coupling in exciton-polaritons**

We engineered a dielectric metasurface, consisting of isosceles triangular holes arranged in a square lattice, carved directly in a polycrystalline methylammonium lead iodide ($MAPbI_3$) film (Methods and *Supplementary Information, Section 1*) coated on a $SiO_2$/Si substrate (**Fig. 1a**). Due to their geometric rotational asymmetry, the isosceles triangles intrinsically disrupt the twofold rotational invariance of the structure[48,49], enabling the mode transition from a symmetry protected BIC with the characteristic vortex-like (V-point) polarization singularity, to a symmetry-broken *q*BIC, with two distinct and opposite spin states, $\sigma^+$and $\sigma^-$, located off the centre ($\Gamma$−point) of the momentum space (*Supplementary Information, Section 2*). Owing to the large oscillator strength of the $MAPbI_3$ exciton, the reduced mode volume of the *q*BIC resonance, and the indistinguishable spatial overlap between them afforded by the monolithic configuration (photonic and excitonic resonances are defined within the same structure), exciton-polaritons are formed and undergo condensation transition above a threshold excitation fluence. The two opposite spin-polarized polariton states carry a half-integer topological charge $\mathrm{p} = \pm\frac{1}{2}$ associated with a $\pi$ rotation of the polarization ellipse, and an integer topological charge $\mathrm{q} = \pm 1$ related to the winding of their phase. This results in circularly polarized condensate emission with opposite handedness and high spin purity, in counter-propagating, momentum-locked directions (**Fig. 1b**).

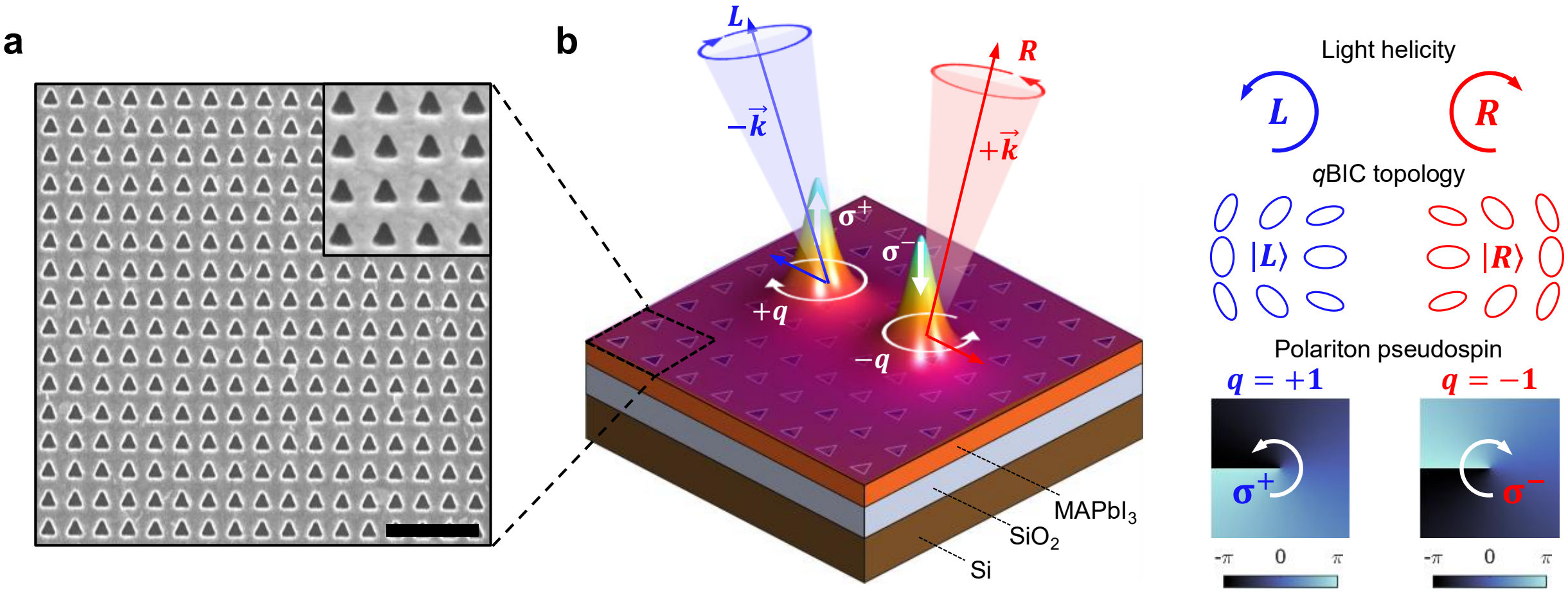

**Figure 1. Spin-orbit coupling of exciton-polaritons from broken inversion symmetry BIC metasurface: (a)** Scanning electron microscope (SEM) image of the monolithic perovskite square lattice metasurface with isosceles triangular holes (scale bar 1 μm). The inset shows a magnified view of the triangular metamolecules. **(b)** Breaking of the in-plane $C_2$ symmetry splits the BIC resonance into a pair of opposite circularly polarized states $\sigma^+$and $\sigma^-$, enabling the formation of a spin-orbit coupled exciton-polariton condensate with half-integer topological charges $p = \pm\frac{1}{2}$ in the polarization domain, and integer charges $q = \pm 1$ in the phase domain.

**Spin-polarized *q*BIC exciton-polariton condensate**

The interaction between the $MAPbI_3$ exciton and the uncoupled *q*BIC can be tailored by tuning the metasurface geometrical parameters to maximize their mutual overlap and enable formation of hybrid polariton states with high spin purity. **Figure 2** shows the emergence of strong-coupling and the transition of the exciton-polariton to a condensate state through simulated and experimental angle-resolved spectra of the metasurface. The energy-momentum maps of numerically calculated transverse-electric (TE) transmittance (**Fig. 2a**) and the experimental photoluminescence (**Fig. 2b**), measured using a Fourier imaging setup (*Supplementary Information, Section 4*), show the appearance of a clear *q*BIC band below the $MAPbI_3$ exciton (1.685 eV at 77 K, dotted green line in **Fig. 2b**). Fitting of the measured bands with a coupled oscillator model between the $MAPbI_3$ exciton and the uncoupled *q*BIC resonance (dotted white line in **Fig. 2b**) captures well the experimental band dispersion and predicts the formation of upper (UP) and lower (LP) exciton-polariton bands (dashed white lines in **Fig. 2b**) with a Rabi splitting of $\hbar\Omega = 102$ meV. Only the lower polariton branch is visible due to the strong absorption of the perovskite at energies above the exciton line. A detailed description of the fitting procedure is provided in *Section 3* of the *Supplementary Information*. By increasing the excitation fluence, polaritons undergo a condensation phase transition, evidenced by the localization of light emission near the Γ-point (**Fig. 2c**). This coincides with the onset of polariton lasing at a threshold $P_{th} \cong 26.9\ \mu J/cm^2$, with characteristic sigmoidal shape of the light-light curve, narrowing of the FWHM and spectral blueshift. The polariton laser shows extended spatial and temporal coherence around $10\ \mu m$ and $500\ fs$ , respectively (*Supplementary Information, Section 4*). The monolithic integration of the metasurface within the perovskite film results in a reduced condensation threshold and increased spatial coherence length compared to what observed in similar polycrystalline films coupled to dielectric metasurfaces[50], owing to the reduced spatial inhomogeneities enabled by the monolithic design. As expected from the in-plane inversion symmetry breaking in the *q*BIC metasurfaces, the polariton bands show a high degree of polarization, with strong dichroism appearing between modes emitted in opposite directions relative to the centre of the Brillouin zone. To map the location of the two points with circular polarization in the momentum space (C-points in the Poincaré sphere), we retrieved the Stokes parameter $S_3$ along the lower polariton branch from eigenfrequency simulations. Starting from large k values, the ellipticity first changes sign and then reduces approaching the Γ-point , becoming circular ($|S_3| = 1$) at $k_x = \pm 0.38\ \mu m^{-1}$ (**Fig. 2d**). The experimental $S_3$ parameters agree well with the trend predicted by numerical simulations, yielding polariton pseudospin purity of $|S_3| \sim 0.6$ below threshold (**Fig. 2e**). We note that, close to the sign inversion region ($k_x \cong \pm 2.2\ \mu m^{-1}$), the apparent energy splitting of the dispersions occurs within the linewidth of the polariton band and does not indicate a real splitting but reflects the increased sensitivity of polarization measurements as the $S_3$ approaches zero. Above the excitation threshold, polariton condensation occurs exactly at the momentum-space locations corresponding to the C-points, and the spin purity increases to $|S_3| \sim 0.85$ (**Fig. 2f**). This provides clear evidence for the emergence of a polariton condensate with well-defined pseudospin directly connected to the topology of the metasurface. The slight residual ellipticity can be attributed to fabrication imperfections.

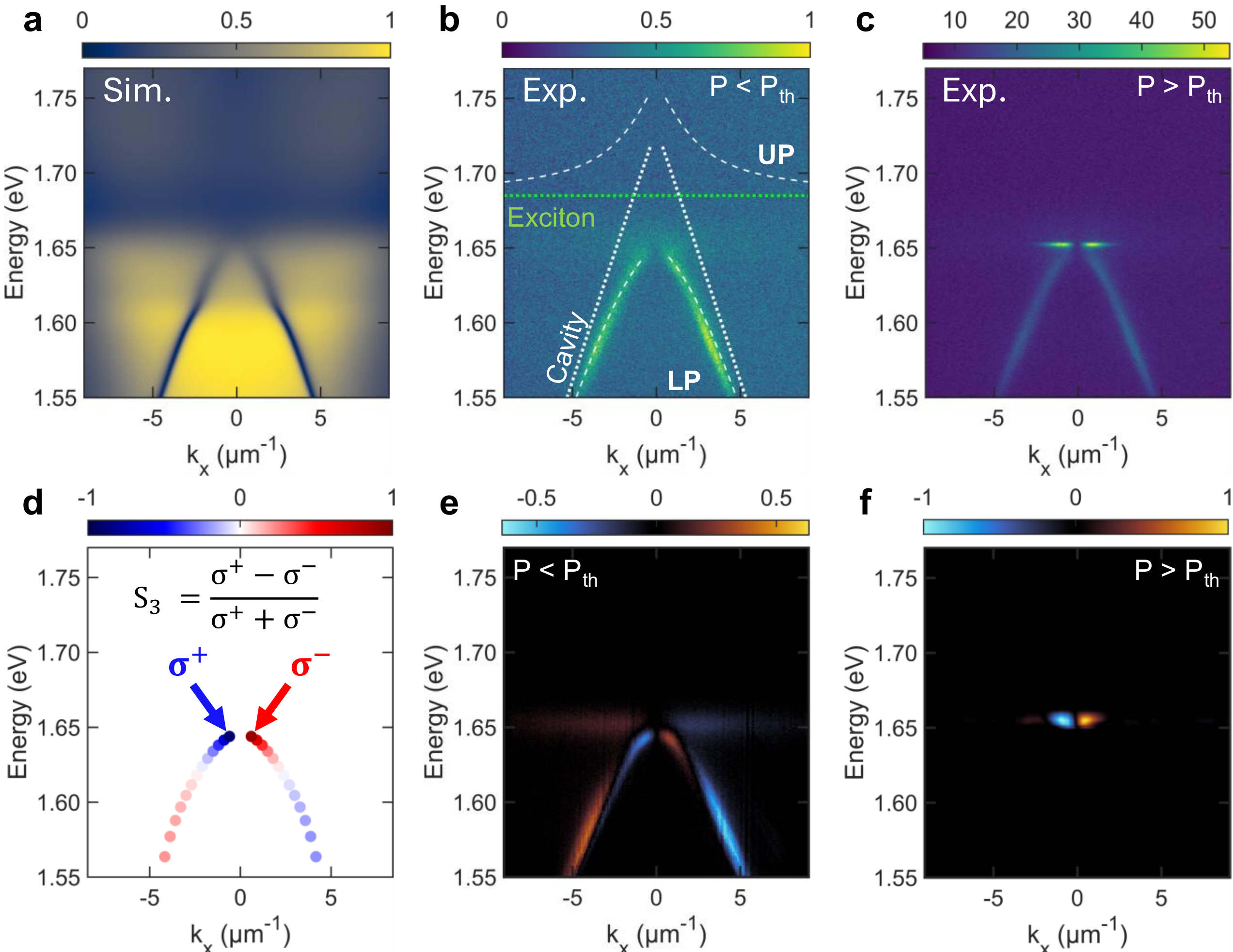

**Figure 2. Characterization of spin-polarized *q*BIC-polariton condensate: (a)** Simulated transmittance and **(b)** measured photoluminescence angle-resolved spectra of the *q*BIC metasurface showing anticrossing and the formation of the lower exciton-polariton branch. The uncoupled cavity resonance (dotted white line) and the $MAPbI_3$ exciton (1.685 eV, dotted green line) in **(b)** are used to fit the polariton bands (dashed white lines) through a coupled oscillator model. **(c)** Above the excitation threshold $P_{th} = 26.9\ \mathrm{uJ/cm^2}$, the system exhibits condensate phase transition with the characteristic BIC emission lobes localized near the center of the reciprocal space. **(d)** Numerical simulations predict spin anisotropy of the band in the momentum space, that it is successfully imprinted to the LP emission below threshold **(e)** and to its condensate phase **(f)**.

To gain a complete picture of the polarization landscape of the condensate and its connection to the underlying metasurface topology, we performed full Stokes polarimetry of the emission, both in reciprocal (**Fig. 3a-d**) and in real (**Fig. 3e-h**) space. The normalized intensity distributions in both domains reveal two symmetric lobes with a pronounced central intensity minimum, where the emission nearly vanishes **(Fig. 3a,e)**, while the Stokes parameters ($S_1$, $S_2$, $S_3$) uncover a highly structured pseudospin texture. Besides the presence of two regions of opposite circular polarization associated with the C-points ($S_3$ parameter in **Fig. 3d,h**), both linear polarization components reveal a smooth yet nontrivial rotation of the basis across both real and reciprocal space ($S_1$ and $S_2$ parameters in **Fig. 3b,f** and **Fig. 3c,g**, respectively). This continuous evolution of the polarization state, accompanied by localized regions of rapid variation, highlights the formation of a complex pseudospin texture directly inherited from the underlying qBIC metasurface. Notably, the polarization distributions observed in real and reciprocal space are in strong agreement, indicating that the mapping of the condensate pseudospin is consistent across the two domains. Furthermore, both measurements are in good quantitative agreement with the numerical simulations of the Stokes parameters reported in the *Supplementary Information, Section 6*.

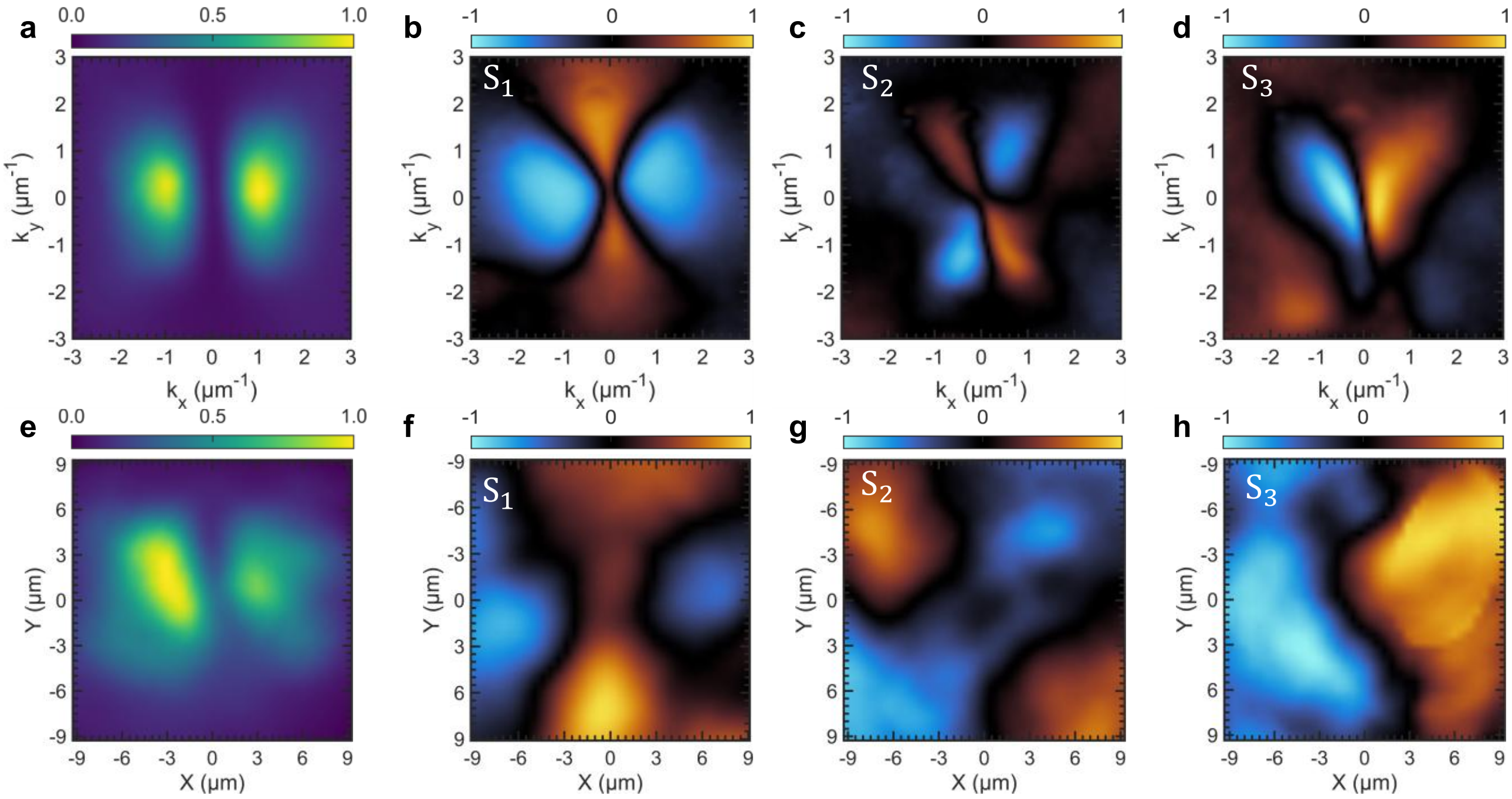

**Figure 3. Stokes polarimetry of the condensate emission: (a–d)** Normalized intensity and corresponding Stokes parameters $S_1$, $S_2$, and $S_3$ of the condensate emission in reciprocal space, revealing the momentum-resolved pseudospin texture. **(e–h)** Real-space normalized intensity and corresponding Stokes parameters, showing the spatial distribution of the condensate polarization.

**Half-vortices in the $q$BIC condensate emission**

Half-vortices in polariton condensates were initially predicted by Y. Rubo[51] and subsequently observed in CdTe/CdMgTe microcavities under non-resonant excitation[52]. A half-vortex is characterized by a phase singularity in only one circular polarization and is described by a two winding numbers $(n, m)$, where $n$ denotes the winding of the polarization angle $\eta$ and $m$ the one of the gauge phase $\phi$, that change simultaneously by $\pm\pi$ [53,54]. In analogy with the spinor formulation introduced in Ref. 51, the condensate field can be described using a two-component spinor in a linear basis $u_L(\mathbf{k})$ uniquely imparted by the $q$BIC topology:

$$u_L(\mathbf{k}) = e^{i\vartheta(\mathbf{k})} \begin{bmatrix} \cos\eta(\mathbf{k}) \\ \sin\eta(\mathbf{k}) \end{bmatrix} \tag{1}$$

Topological defects are identified by examining the winding of the angles $\vartheta$ and $\eta$ upon encircling singular points in momentum space. Specifically, a closed loop in k-space may induce rotations of the order parameters $\eta(\mathbf{k}) \rightarrow \eta(\mathbf{k}) + 2\pi n$ and $\vartheta(\mathbf{k}) \rightarrow \vartheta(\mathbf{k}) + 2\pi m$, where $n$ and $m$ can be integer or half-integer and subject to the constraint $(n + m) \in \mathbb{Z}$. To elucidate the nature of these topological configurations, it is convenient to transform the spinor from linear (1) to circular basis via a $U(2)$ transformation[55]:

$$u_C(\mathbf{k}) = \begin{bmatrix} 1 & i \\ 1 & -i \end{bmatrix} e^{i\vartheta(\mathbf{k})} \begin{bmatrix} \cos\eta(\mathbf{k}) \\ \sin\eta(\mathbf{k}) \end{bmatrix} = \frac{1}{\sqrt{2}} \begin{bmatrix} e^{i(\vartheta(\mathbf{k})+\eta(\mathbf{k}))} \\ e^{i(\vartheta(\mathbf{k})-\eta(\mathbf{k}))} \end{bmatrix} \tag{2}$$

In this basis, the two spinorial components acquire phases proportional to $\vartheta(\mathbf{k}) \pm \eta(\mathbf{k})$. As a result, a simultaneous $\pi$ winding of $\eta$ $\left(n = \frac{1}{2}\right)$ and $\vartheta$ $\left(m = \frac{1}{2}\right)$ leads to a full $2\pi$ phase winding of one of the two circular polarization components, while the other remains constant. The non-trivial windings encoded in momentum space is transferred to the real space distribution of the condensate emission and thus can be directly investigated via Michelson interferometry, enabling reconstruction of the wavefront in a selectively chosen polarization basis. In real space, the condensate emission forms two lobes with opposite circular polarization, separated by a dark region characterized by linear polarization, corresponding to the long lifetime $q$BIC dark state (**Figure 4**). Vortex topological charge summation can be achieved by interfering the portion of the emission with the same spin state, $\sigma^+$ (**Fig. 4a)** or $\sigma^-$ (**Fig. 4d)**. This is obtained by selecting the spin state through appropriately aligned quarter waveplate and polarizer, splitting the signal in two arms, and allowing the interference between points $\mathbf{r_1} = (\mathrm{r_x}, \mathrm{r_y})$ of the original signal L1 (R1), and points $\mathbf{r_2} = (\mathrm{r_x}, -\mathrm{r_y})$ of the retroreflected wavefront L2 (R2). The interference fringes are generated by the phase difference between the two beams and the presence of bifurcations, seen in **Figs. 4b, e**, indicate the existence of vortex phase singularities. The maps of the phase difference between signals on the two arms, $\Delta\phi = \varphi_1 - \varphi_2$, confirm the presence of spatially-separated phase vortices carrying opposite topological charges, with the $\sigma^+$ component exhibiting a positive charge, $q = +1$ (**Fig. 4c**), and the $\sigma^-$ component a negative charge, $q = -1$ (**Fig. 4f**). The two vortices seen in each phase maps belong to the signal (L1 or R1) and retroreflected (L2 or R2) beam and carry the same charge because of the mirror flipping introduced by the interferometer arms. Notably, at the spatial positions where a half-vortex is present in one spin component, the complementary spin component displays a flat phase profile, consistent with the absence of phase singularities. As a further validation of the interferometric measurements, we superimposed the two lobes carrying opposite topological charges by mirroring the retroreflected image as $\mathbf{r_2} = (-\mathrm{r_x}, \mathrm{r_y})$, as illustrated in **Fig. 4g**. In this configuration, the $\sigma^+$ and $\sigma^-$ fields are overlapped with opposite helicity, the resulting fringe pattern (**Fig. 4h**) is free of dislocations, and the vortices are absent in the retrieved phase map (**Fig. 4i**). The disappearance of the fork-like bifurcations indicates that the condensate hosts a V-AV pair generated by the inherent spin-orbit interaction of the metasurface and confirms that cavity engineering is a viable route to control topological charges in polariton condensates. Additional information on the vortex core size and spin-polarized emission is provided in the *Supplementary Information, Section 8*.

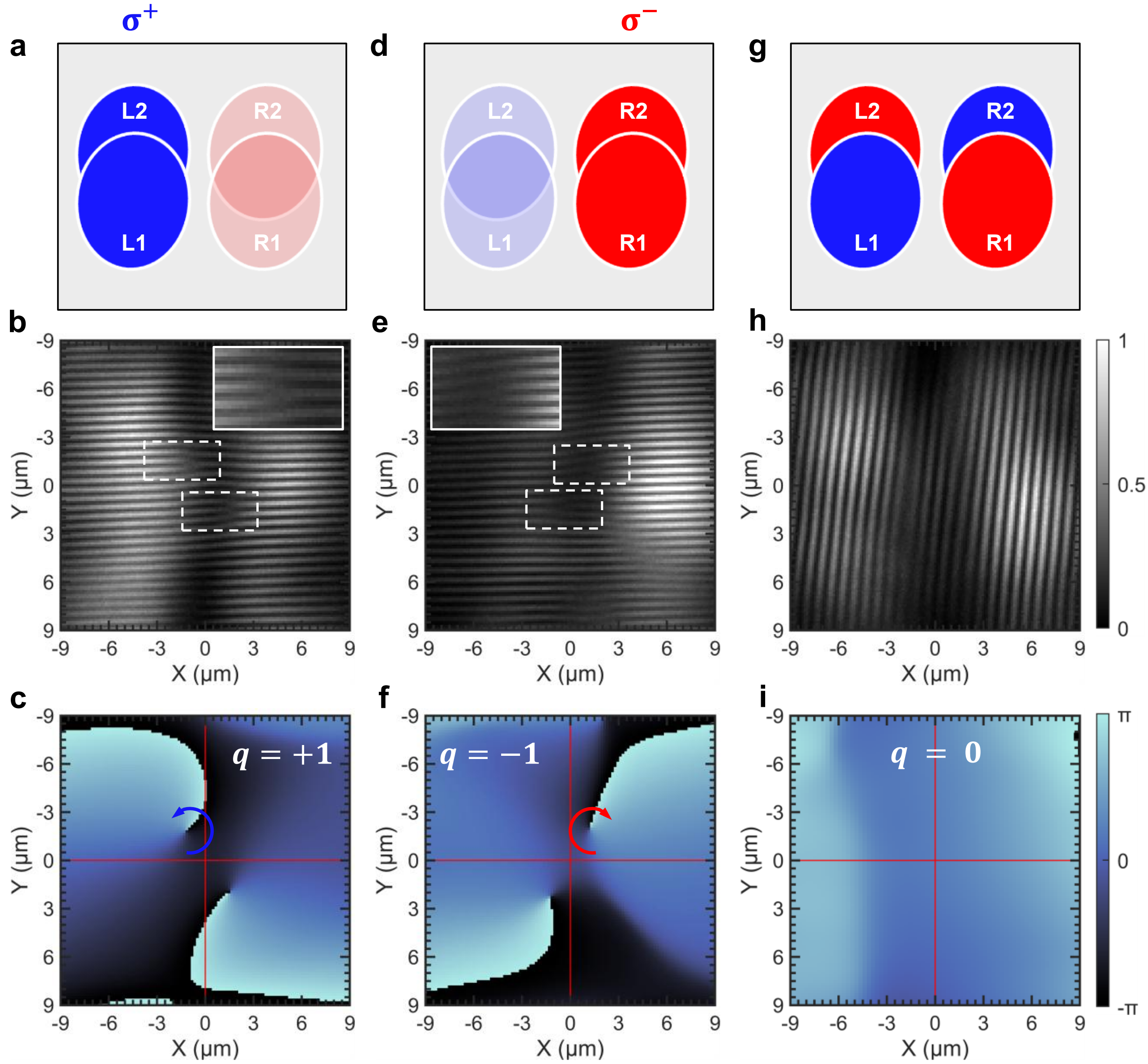

**Figure 4. Real space half-vortices observation in qBIC polariton condensate: (a)** Schematic of the Michelson interferometer configuration, where the lobe with left circular polarization L1 is interfered with its retroreflected image L2. **(b)** The horizontal interference fringes present characteristic bifurcations patterns (encircled by dashed lines; the inset shows a magnified view of one of the bifurcations) and **(c)** the retrieved phase map reveals the presence of vortices carrying a topological charge $q = +1$. **(d)** Same interferometer configuration of **(a)** for the lobe with the right circular polarization R1 interfered with R2, showing opposite bifurcation patterns in the interference fringes **(e)** and vortices with topological charge $q = -1$ **(f)**. **(g)** Spatial overlap of emission lobes with opposite polarization, achieved rotating the retroreflector by 90° resulting in **(h)** vertical interference fringes without any bifurcations and **(i)** complete disappearance of the vortices.

**Strings, domains, and topological interaction in a $q$BIC metasurface condensate**

Spin plays a major role in governing the interactions between condensate polaritons. In general, polaritons with spin of the same sign repel each other, while those with opposite sign attract[56]. These attractive and repulsive interactions are directly reflected into vortices dynamics, that undergo repulsion and attraction according to the sign of the topological charges they carry[57]. Time-resolved dynamics studies revealed that vortices are not spatially pinned in time, but their positions and angular orientations follow complex uncontrollable trajectories that depend on the nature of the vortices, the initial injection conditions, and the condensate particle density[58]. Thus, it is of particular interest to explore whether topological bounds emerging from the cavity geometry can impose deterministic constraints on vortex motion, effectively guiding and stabilizing their trajectories.

Strings attached to vortices are well known high-order composite topological objects in multicomponent superfluids[59], where they appear as solitonic regions connecting topological defects. An example is the Alice string discussed in superfluid phase $^3$He-A, that corresponds to a region where the internal order parameter of the system (i.e. orbital and spin vectors) undergoes a π rotation and mediates effective interactions between defects[60].

Analogously, within the polariton condensate the string is defined as the line across which the polarization angle η spans an arc of length $\pm\pi$ in the Poincaré sphere,[61] (i.e., when crossing the line in the locally perpendicular direction), that corresponds to $\pm\frac{\pi}{2}$ spatial flip of the polarization ellipse. To investigate the presence of these features in our platform, we derived the cosine of the polarization angle, $\cos(\eta)$ with $\eta = \frac{1}{2}\tan^{-1} S_2/S_1$, in both reciprocal (**Fig. 5a**) and real space (**Fig. 5b**). Two symmetric regions exhibit a rapid variation of $\cos(\eta)$, following an approximately parabolic profile (dotted black lines) that passes through the vortex cores. Across the direction perpendicular to these curves, cos(η) rapidly changes from 0 to 1, corresponding to an exact π shift of the polarization angle. These features therefore correspond to the polarization strings attached to the vortex cores within the condensate, delineating the boundaries of a domain wall at the center of the condensate. These act as spatial constraints, limiting the vortices' degrees of freedom and preventing the overlap and annihilation of the vortex-antivortex pair.

The emergence of a domain wall is consistent with Stokes polarimetry (**Fig. 3b,f**) , where the polarization at the center of the Brillouin zone is constrained to linear by the residual symmetry of the system, defining a region where vortices cannot form. An abrupt π-phase discontinuity of the linear polarization at the Γ point (**Fig. 5c**) directly identifies the centre of the domain wall and determines the central nodal line in the emission intensity (**Fig. 3a,e**).

From a complementary perspective, the domain wall corresponds to the boundary between regions of opposite Berry curvature, where the curvature changes sign and vanishes (**Fig. S15**). This boundary partitions the Brillouin zone into regions of opposite geometric flux, and enforces a constraint on the pseudospin, which is forced to lie on the equator of the Poincaré sphere, corresponding to a purely linear polarization state. Owing to the residual symmetry of the system, this condition is pinned at the center of the Brillouin zone, stabilizing the domain wall at this location. A detailed experimental analysis of the Berry curvature is provided in the *Supplementary Information, Section 10*.

In light of the above observations, we interfered the entire condensate emission beam with its mirror image to simultaneously visualize both vortices and track their spatial positions under increasing excitation fluence (see *Supplementary Information, Section 9*). As shown in **Fig. 5d**, the vortex-antivortex positions evolve with increasing fluences along well-defined trajectories (dotted black lines), that corresponds to the boundaries of the domain. None of these trajectories crosses the center of the emission area, reflecting the existence of a central domain that prevents their mutual annihilation. In this framework, the spatial displacement of the vortices as function of excitation density is governed by the metasurface geometry, with the strings and the domain wall defining topological pathways that guide their spatial arrangement. Notably, as the excitation density increases, the vortices are observed at progressively larger separations, that is consistent with an increase in polariton population, which shifts their equilibrium positions, leading to a variation of the inter-vortex distances.

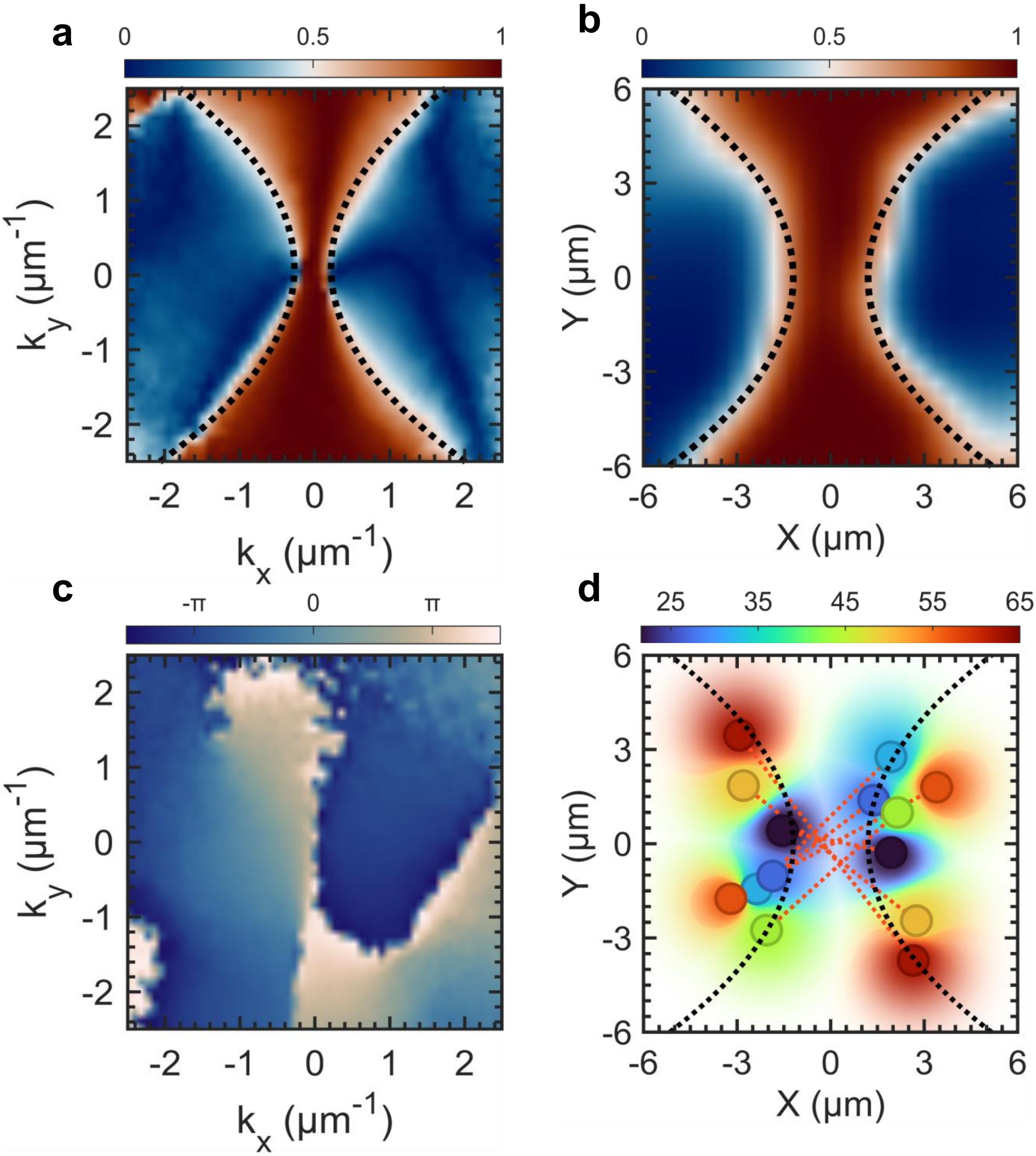

**Figure 5. Topological strings, domain wall, and vortices trajectory: (a)** Reciprocal and **(b)** real space distribution of the cosine of the angle of polarization, $cos(\eta)$. The dotted black lines highlight the emerging strings within the condensate emission. The polarization angle flips rapidly along the direction perpendicular to the strings. **(c)** Experimental phase of linear polarization highlighting a nodal line in the center of the Brillouin zone characterized by a $\pi$ discontinuity and reflecting the presence of a domain wall. **(d)** Spatial evolution of the vortex pair positions as a function of increasing excitation fluence. The vortices follow well-defined trajectories constrained by the domain wall boundaries (highlighted by the dotted black lines). The underlying topology also prevents both boundary crossing and vortex-antivortex annihilation. The colormap represents the excitation fluence and is interpolated in space to guide the visualization of vortices positions, while the dotted orange lines connect each the vortex pair.

## Conclusion

We demonstrated the generation of a half-vortices within a polariton condensate through the topology and high mode confinement of a $q$BIC monolithic perovskite metasurface with broken in-plane inversion symmetry. The $q$BIC encodes spin-orbit coupling in the cavity modes, which is inherited by the strongly coupled exciton-polaritons and deterministically imprints a momentum-locked spin texture onto the condensate. The metasurface topology not only seeds a spatially separated half vortex-antivortex pair with opposite charges and locations but also governs its spatial configuration and density dependent motion through the emergence of strings and a central domain wall. These composite topological objects define a rich topological landscape that constrains the accessible pathways of the vortices, enabling geometry-driven control over their positions while preventing mutual annihilation.

This constitutes both a significant paradigm shift from conventional methods used to access the polariton pseudospin, which require external artificial fields, and demonstrate the possibility to generate and manipulate topological excitations solely through geometry, even in disordered polycrystalline films, thanks to the robustness against structural imperfections.

Overall, the present results establish topological metasurfaces as a versatile platform for exploring advanced topological phenomena in quantum fluids, including spin Hall[62] and meron Hall[63] currents, stable vortex crystals[64], and vortex molecules[65], potentially enabling their application in information encoding schemes[66], and the realization of vortex all-optical logic operations[67].

## Methods

### Metasurface design

The quasi-bound state in the continuum supporting metasurface was designed via Finite Difference Time Domain (FDTD) and Rigorous Coupled-Wave Analysis (RCWA) suit of the commercial software Ansys Lumerical. The adopted triangular geometry supports an out-of-plane magnetic $H_z$ quadrupole resonance with an asymmetric distribution in the x-y plane dictated by the broken in-plane inversion symmetry of the design (*Supplementary Information*, *Section 2*). COMSOL Multiphysics eigenfrequency solver has also utilized to calculate the quality factor and the complex electric field of the resonance at multiple angles.

### Polycrystalline $MAPbI_3$ film deposition and metasurface fabrication

$MAPbI_3$ films were fabricated from a 0.5 M precursor solution of $CH_3NH_3I$ (Dyesol, Greatcell) and $PbI_2$ (99.99%, Sigma-Aldrich) with a molar ratio 1:1, in anhydrous dimethylformamide (DMF, Sigma-Aldrich). The solution was magnetically stirred over night at 373 K in $N_2$ inert atmosphere glovebox and then filtered by a polyvinylidene fluoride (PVDF) syringe filter (0.45 μm) before spin-coating. Prior to deposition, the substrates were cleaned with ultrasonication for 5 min in acetone and isopropanol, respectively. Subsequently, substrates were dried with nitrogen, followed by 20 minutes ozone surface treatment. The perovskite precursor solution was spin-coated onto the $Si/SiO_2$ substrates with a speed of 4000 rpm for 35 s using the toluene anti-solvent deposition method, with the anti-solvent being drop-cast on the substrates 6 s after spinning began. The film was then annealed at 373K for 10 minutes, yielding a 131 nm thickness, as measured by atomic force microscopy (*Supplementary Information*, *Section 1*). The perovskite metasurface fabrication was conducted by focused ion beam (FIB) lithography process. A 50 x 50 $\mu m^2$ triangular air holes array was patterned using the Helios 600 NanoLab, FEI system. The ion beam current was controlled around ~1.7 pA for suitable spot size of etching.

**Angle-resolved spectroscopy**

Angle-resolved reflectance (ARR) and photoluminescence (ARPL) measurements were performed with a custom-built micro-spectrometer setup consisting of an inverted optical microscope (Nikon Ti-u, 50x objective, NA=0.556), a spectrograph (Andor SR-303i with a 300 lines/mm grating), and a charged-coupled detector (CCD, Andor iDus 420). A series of lenses along the beam path between the microscope and the spectrograph projects the back focal plane (BFP) of the collection objective on the slit of the spectrograph, allowing the collection of angular information within bounds defined by $k_x/k_0$=NA=0.556. The sample was excited by a 400 nm femtosecond pulsed laser (Coherent) with 1 KHz repetition rate and beam diameter of 17 μm.

**Stokes polarimetry**

The Stokes parameters were retrieved using a standard over-complete polarimetry scheme based on six intensity measurements covering completely the polarization basis (H, V, D, A, L, R). This approach introduces internal redundancy, enabling independent determination of each Stokes component and mitigating the effects of polarization-dependent detection efficiency and imperfections in the optical elements.[68,69] The linear polarization was obtained by rotating the fast axis of a half-wave plate with respect to a fixed linear polarizer, while the circular components were measured rotating the fast axis of a quarter-wave plate with respect to a fixed linear polarizer. Since the analysis does not rely on the absolute value of the measured polarization but on the topology of the field, the identification of polarization singularities and vortex charges is insensitive to possible systematic uncertainties in the polarimetric calibration.

**Measurements of spatial and temporal coherence**

Spatial and temporal coherence of the condensate are inferred by the interference pattern formed in a standard Michelson interferometer (see *Supplementary Information*, *Section 4*), where the sample emission is split into two arms by a 50/50 beamsplitter. The two arms are mounted on translating stages to spatially and temporally overlap the reflect images with a retroreflector and a mirror to form interference fringes. The retroreflector is mounted on a coarse alignment stage, while the mirror is mounted on a precision micrometre actuator with 1 μm resolution (model SM1ZA). The two arms are then recombined on a CMOS camera (model CS165MU).

**Author contributions**

A.Z. G.A. and C.S. conceived the idea. A.Z. carried out design optimisation and fabrication of the perovskite metasurfaces and conducted spectroscopy measurements and data analysis with the help of M.M and G.A. A.Z. and N.M.C. performed all numerical simulations. A.Z. drafted the manuscript and all authors contributed to editing and review. C.S. and G.A. supervised the work.

**Acknowledgements**

We are indebted to D. Gerace, T.C.H. Liew and Y. Rubo for the insightful discussions and suggestions they provided with regard this work. Research was supported by the Singapore Ministry of Education (Grant no. MOE-T2EP50222-0015) and the National Research Foundation through the National Centre for Advanced Integrated Photonics (Grant no. NRF-MSG-2023-0002). Y.S. acknowledges support from the Singapore Ministry of Education (RG157/23 and RT11/23), the Singapore Agency for Science, Technology and Research (A*STAR) (M24N7c0080 and 256I9013), and the Nanyang Assistant Professorship Start Up grant.